\documentclass[a4paper]{jpconf}
\usepackage{graphicx}
\usepackage{amssymb}

\begin{document}
\title{Magnetic order in the filled skutterudites {\it R}Pt$_4$Ge$_{12}$ ({\it R} = Nd, Eu)}

\author{M Nicklas,  R Gumeniuk, W Schnelle, H Rosner, A Leithe-Jasper,\\ F Steglich and Yu Grin}
\address{Max\,Planck\,Institute\,for\,Chemical\,Physics\,of\,Solids,
N\"{o}thnitzer Str.\,40, 01187\,Dresden, Germany} \ead{nicklas@cpfs.mpg.de (Michael Nicklas)}

\begin{abstract}
Rare-earth metal filled skutterudites {\it R}Pt$_4$Ge$_{12}$ with {\it R}=La-Nd, and Eu exhibit a
variety of different ground states, e.g., conventional and unconventional super\-conductivity in
LaPt$_4$Ge$_{12}$ and PrPt$_4$Ge$_{12}$, respectively, and intermediate valence behavior in
CePt$_4$Ge$_{12}$. %, and magnetic order in NdPt$_4$Ge$_{12}$ and EuPt$_4$Ge$_{12}$.
In this work we investigate the magnetic state of NdPt$_4$Ge$_{12}$ and EuPt$_4$Ge$_{12}$ by specific
heat, dc-susceptibility and magnetization. NdPt$_4$Ge$_{12}$ shows two magnetic phase transitions at
$T_{N1}=0.67$~K and $T_{N2}=0.58$~K, while EuPt$_4$Ge$_{12}$ displays a complex magnetic phase
diagram below the magnetic ordering temperature of 1.78~K. The specific heat indicates that in
NdPt$_4$Ge$_{12}$ the crystalline electric field (CEF) ground state of the Nd$^{3+}$ is a quartet and
that, as expected, in EuPt$_4$Ge$_{12}$ the Eu$^{2+}$ state is fully degenerate.
\end{abstract}

\section{Introduction}
Filled skutterudites have become a topic of considerable interest with respect to basic and applied
solid state sciences. This is particularly due to a variety of physical properties which can be
intimately related to the underlying structural chemistry
\cite{Sales1996,Uher2001,Chen2003,Uher2008,Sales2007}. Their stoichiometry can be rationalized with
the chemical formula {\it MT}$_4${\it X}$_{12}$, with {\it M} being an electropositive element like
alkali, alkaline-earth, early rare-earth, actinide, or thallium metal, {\it T} standing for a
transition metal of the iron- or cobalt-group, and {\it X} representing a pnictogen element as they
are phosphorus, arsenic, or antimony. They all crystallize with the cubic LaFe$_4$P$_{12}$
\cite{Jeitschko1977} structure where cation {\it M} stabilizes the {\it T}$_4${\it X}$_{12}$ host
structure. Recently it has been shown by us and others \cite{Bauer2007,Gumeniuk2008} that the
transition metal is not restricted to the iron or cobalt group, but can also be the noble metal
platinum, which together with germanium acts as the framework forming elements stabilized by the
alkaline-earth metals Sr and Ba. Moreover, we have discovered a whole new family of rare-earth metal
({\it R}) based filled skutterudites {\it R}Pt$_4$Ge$_{12}$ with {\it R} = La-Nd, and Eu
\cite{Gumeniuk2008}. In turn, this series of compounds has been extended to actinide based {\it
A}Pt$_4$Ge$_{12}$ ({\it A}= Th, U) phases \cite{Kaczorowski2008,Bauer2008}. {\it M}Pt$_4$Ge$_{12}$
compounds ({\it M} = Sr, Ba, La, Pr) are superconductors with $T_c$ up to 8.3 K \cite{Gumeniuk2008}.
PrPt$_4$Ge$_{12}$ is an unconventional superconductor with nodes in the superconducting energy gap
\cite{Maisuradze2009}. Experimental and theoretical analysis of the electronic structure and chemical
bonding revealed deep-lying Pt 5{\it d} states which only partially form covalent bands with the Ge
4{\it p} electrons. Consequently, the states at the Fermi level, which are relevant for
superconductivity of these compounds, can be assigned to originate mainly from Ge 4{\it p} electrons
\cite{Rosner2009}. Here, we report on the low-temperature magnetic properties of {\it
R}Pt$_4$Ge$_{12}$ ({\it R} = Nd, Eu) studied by specific heat, dc-susceptibility, and magnetization
experiments.

\section{Experimental Details}
Samples were synthesized as described elsewhere by arc-melting from the elements followed by
annealing procedures \cite{Gumeniuk2008}. We carried out low temperature ($0.35{\rm K}\leq T\leq
10{\rm K}$) specific-heat ($C$) experiments using the $^3$He option of a PPMS (Quantum Design). The
magnetization ($M$) in the temperature range $0.48{\rm K}\leq T\leq 1.9{\rm K}$ was measured using a
SQUID magnetometer (MPMS, Quantum Design) equipped with a $^3$He option (iQuantum). X-ray absorption
spectroscopy at the Eu $L_{\rm III}$ edge (6977~eV) was used to determine valence of this rare-earth
metal in the structures of new filled skutterudite.

\section{Experimental Results}
\subsection{EuPt$_4$Ge$_{12}$}

\begin{figure}[t!]
\begin{minipage}{18pc}
\includegraphics[width=18pc]{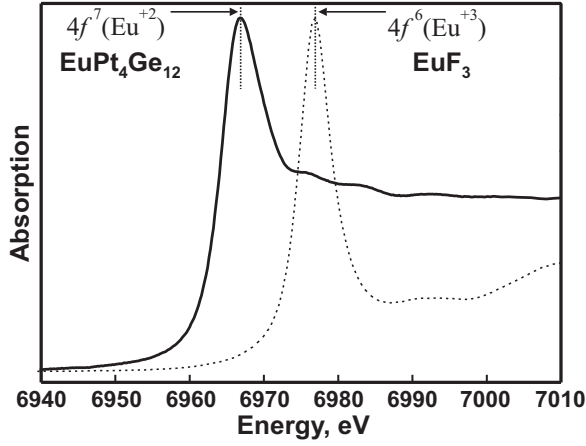}
\end{minipage}\hspace{2pc}
\begin{minipage}{18pc}\vspace{-5pc}\caption{\label{XANES} XAS of EuPt$_4$Ge$_{12}$ and EuF$_3$ at the Eu $L_{\rm III}$ edge.
The XAS were recorded at room temperature at the EXAFS beamline A1 of HASYLAB at DESY using the
four-crystal mode of the Si(111) monochromator and determined in transmission mode using powdered
samples ($\sim$10~mg) which were diluted with B$_4$C and embedded in paraffin wax.}
\end{minipage}
\end{figure}
\begin{figure}[b!]
\begin{minipage}{18pc}
\includegraphics[width=18pc]{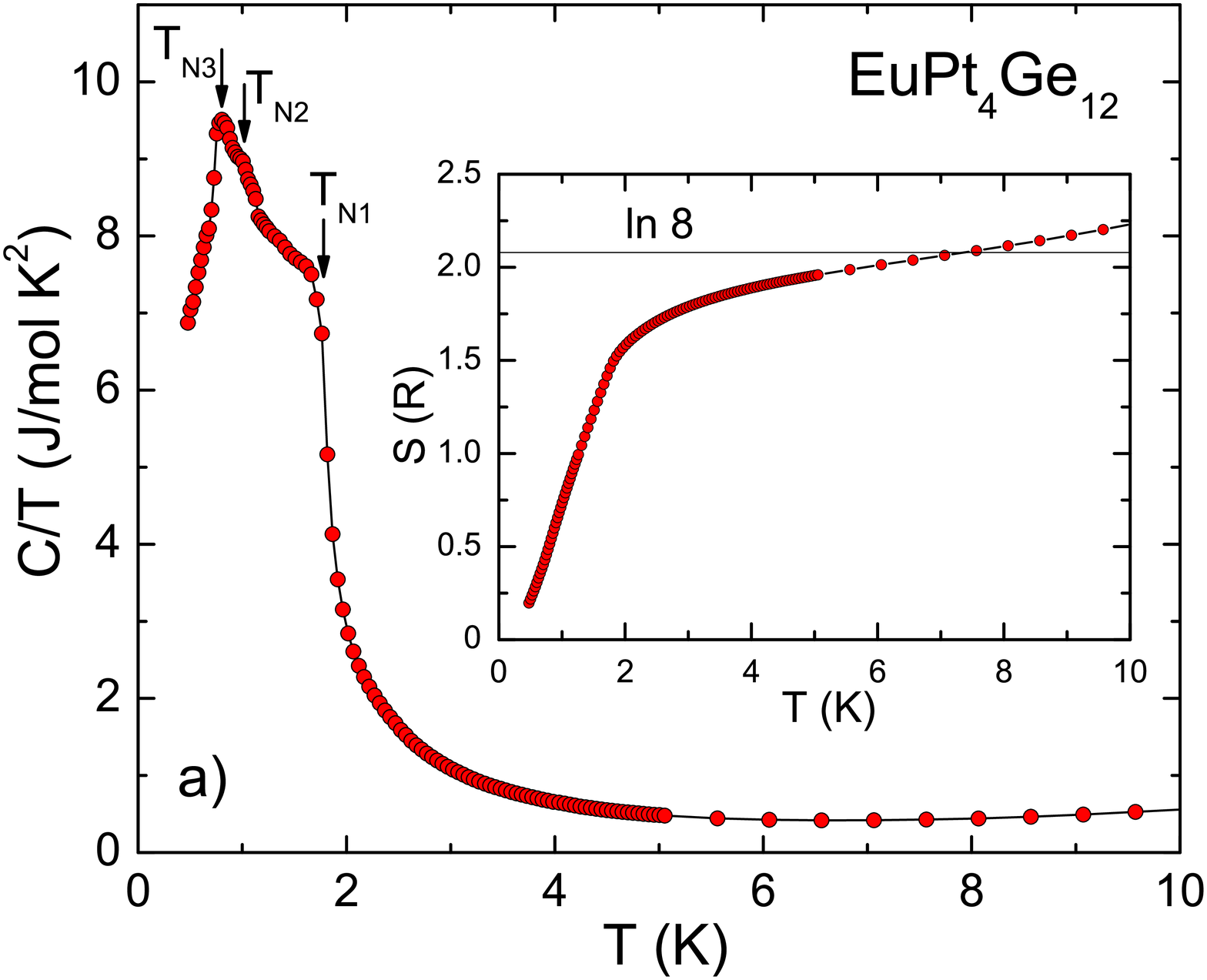}
\end{minipage}\hspace{2pc}%
\begin{minipage}{18pc}
\includegraphics[width=18pc]{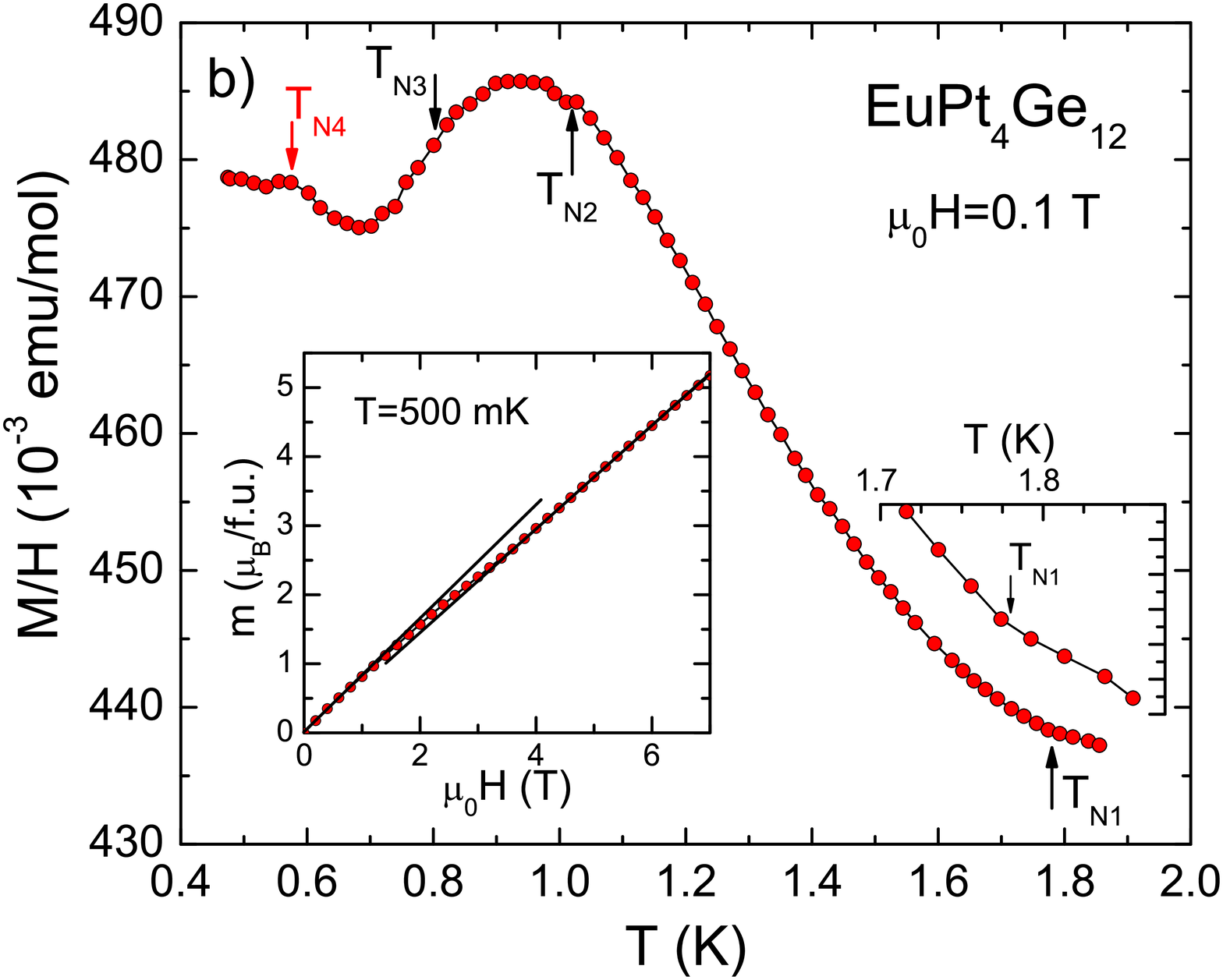}
\end{minipage}\hspace{2pc}%
\caption{\label{Eu} Specific heat as $C(T)/T$ (a) and dc-susceptibility $M(T)/H$ (b) of
EuPt$_4$Ge$_{12}$. The arrows in (a) indicate the phase transitions. The inset of (a) shows the
entropy $S(T)$. The black arrows in (b) correspond to the transition temperatures from specific heat,
while the red arrow indicates an additional anomaly in $M(T)/H$. The inset of (b) presents $m(H)$ at
$T\approx500$~mK.}
\end{figure}

The X-ray absorption spectrum (XAS) for EuPt$_4$Ge$_{12}$ shown in Fig.~\ref{XANES} indicates that Eu
is in the magnetic $4f^7$ (Eu$^{2+}$) configuration. No admixture of the $4f^6$ state is visible in
the data. In Eu$^{2+}$ only the spin component contributes to the total angular momentum of the
ground multiplet $^8$S$_{7/2}$ and, therefore, distinct crystalline electric field effects are
expected to be absent. Accordingly, a fully degenerated $J=7/2$ ground state multiplet is
anticipated. The specific heat, $C(T)/T$, of EuPt$_4$Ge$_{12}$ is displayed in Fig.~\ref{Eu}a. An
analysis of $C(T)/T$ finds -as expected- a magnetic entropy of $R\ln8$ (inset of Fig.~\ref{Eu}a) at
about 7~K.

 $C(T)/T$ exhibits several anomalies at low temperatures. In contrast to previous works,
reporting only one antiferromagnetic transition around $T_N\approx1.7$, we can clearly identify at
least three different anomalies, the first at $T_{N1}=1.78$~K, at a slightly higher temperature than
previously reported \cite{Gumeniuk2008,Grytsiv2008}, and at $T_{N2}=1.03$~K and $T_{N3}=0.8$~K. To
further elucidate the nature of the phase transitions we conducted dc-susceptibility ($M(T)/T$)
measurements. The general shape of $M(T)/H$ is dominated by a strong increase upon decreasing
temperature and a pronounced maximum just below 1~K. Below the maximum $M(T)/H$ decreases again and
saturates toward lower temperatures. This overall shape is typical for antiferromagnetic ordering
with a N\'{e}el temperature slightly below temperature where the maximum in $M(T)/H$ is observed.
However, $C(T)$ exhibits a strong anomaly at a nearly twice as large temperature $T_{N1}=1.78$~K.
Surprisingly, the transition at $T_{N1}$ indicated by an arrow in Fig.~\ref{Eu}b, is hardly visible
in $M(T)/H$. Only in a magnification (right inset of Fig.~\ref{Eu}b) a kink is visible at $T_{N1}$
hinting at a magnetic character of the transition. This transition has been also  previously
identified in resistivity data, while no indications for additional phase transitions at lower
temperatures have been reported \cite{Grytsiv2008}. At $T_{N2}$ a small but clearly visible feature
is evident in both, $C(T)/T$ and $M(T)/H$, while the most pronounced feature in $M(T)/H$ can be
associated with $T_{N3}$. At $T_{N3}$ $C(T)/T$ shows a well developed peak. In $M(T)/H$ one
additional phase transition can be recognized at $T_{N4}$. On a closer look a change in slope is
visible in $C(T)/T$ at this temperature. The magnetization, $M(H)$, at $T\approx0.5$~K is
proportional to the magnetic field in the whole measurement range up to 7~T. We do not observe any
tendency toward saturation since only $\approx5{\rm~\mu_B}$ are attained at our highest field.  Only
at about 2.5~T a tiny step hints at a subtle reorientation of the magnetic moments. Our results
indicate complex magnetic ordering phenomena in EuPt$_4$Ge$_{12}$ which ask for further
investigations.

\subsection{NdPt$_4$Ge$_{12}$}

\begin{figure}[t!]
\begin{minipage}{18pc}
\includegraphics[width=18pc]{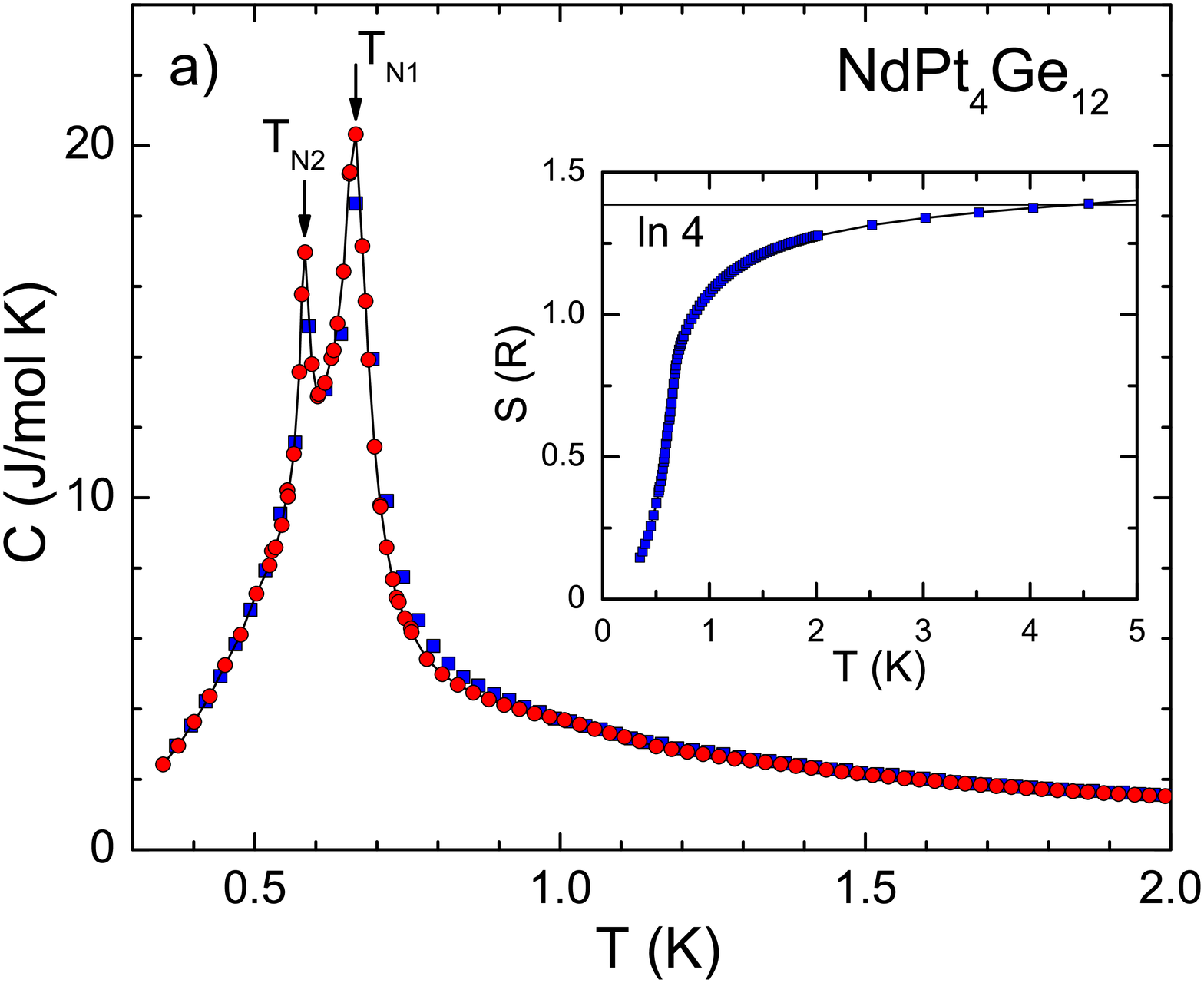}
\end{minipage}\hspace{2pc}%
\begin{minipage}{18pc}
\includegraphics[width=18pc]{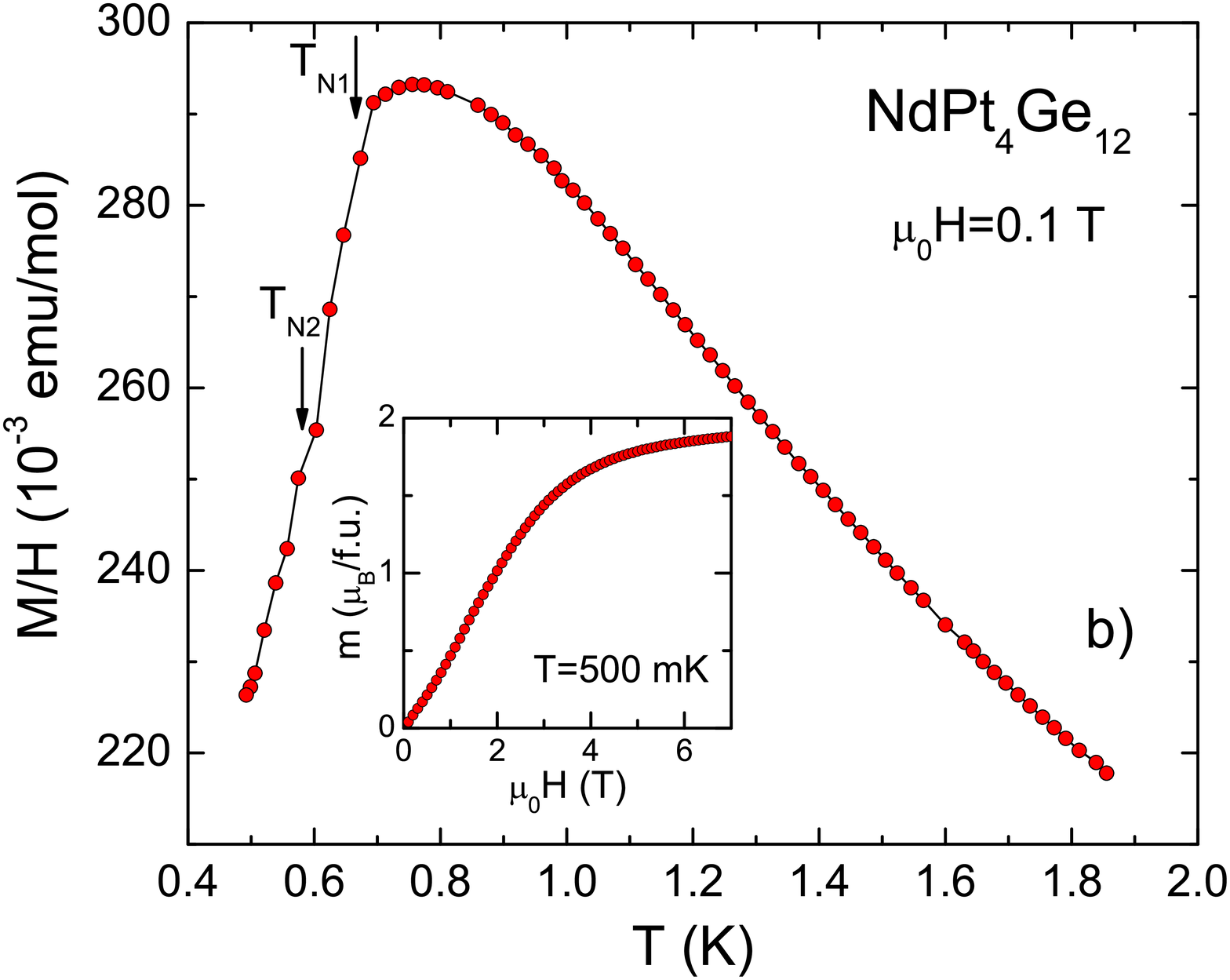}
\end{minipage}\hspace{2pc}%
\caption{\label{Nd} Specific heat as $C(T)/T$ (a) and dc-susceptibility $M(T)/H$ (b) of
NdPt$_4$Ge$_{12}$. For $C(T)/T$ the results from two different measurements are presented. The inset
of (a) shows the entropy $S(T)$ determined form one of the $C(T)$ experiments. The inset of (b)
displays $m(H)$ at $T\approx500$~mK. The arrows in both panels correspond to the transition
temperatures determined from specific heat.}
\end{figure}

The specific heat, $C(T)/T$, of NdPt$_4$Ge$_{12}$ displayed in Fig.~\ref{Nd}a shows two distinct
peaks at $T_{N1}=0.67$~K and $T_{N2}=0.58$~K indicating two magnetic phase transitions. A previous
resistivity study reported a rapid decrease of $\rho(T)$ below $\sim$0.7~K \cite{Toda2008}, which is
in good agreement with $T_{N1}$. However, like in the case of EuPt$_4$Ge$_{12}$, no hint at the
second phase transition has been observed in $\rho(T)$ \cite{Toda2008}. The dc-susceptibility,
$M(T)/T$, shows the typical temperature dependence of an antiferromagnetic material (see
Fig.~\ref{Nd}): with decreasing temperature $M(T)/T$ first continuously increases, exhibiting a
maximum and then sharply drops on further decreasing temperature. As usually observed in an
antiferromagnet the transition temperature ($T_{N1}$) coincides with the onset temperature of the
drop in the susceptibility. At the second phase transition at $T_{N2}$ only a small kink, possibly
indicating a reorientation of the magnetic moments, is evident in $M(T)/T$. The magnetization at
$T\approx0.5$~K deviates from a linear behavior above 4~T, but no saturation is reached up to 7~T. At
7~T a magnetic moment $m=1.88{\rm~\mu_B /f.u.}$ is attained.

An analysis of the magnetic entropy points at a quartet ground state in the CEF level scheme. Already
at $\sim$4~K the complete entropy corresponding to a quartet ground state, $R\ln4$, is recovered, as
displayed in the inset of Fig.~\ref{Nd}a. In a CEF analysis of susceptibility and magnetization data
of NdPt$_4$Ge$_{12}$ indeed a $\Gamma^{(2)}_{67}$ quartet ground state has been favored against a
$\Gamma_{5}$ doublet \cite{Toda2008}. In addition to our entropy analysis also the observed magnetic
moment at 7~T and 0.5~K, $m=1.88{\rm~\mu_B /f.u.}$, is significantly larger compared to the expected
saturation moment in case of a $\Gamma_{5}$ doublet of $1.33{\rm~\mu_B /f.u.}$. Therefore, our data
strongly supports the proposed $\Gamma^{(2)}_{67}$ quartet ground state \cite{Toda2008}.

\section{Summary}

In summary, EuPt$_4$Ge$_{12}$ and NdPt$_4$Ge$_{12}$ order magnetically below 1.78~K and 0.67~K,
respectively. In EuPt$_4$Ge$_{12}$, Eu is in a stable magnetic Eu$^{2+}$ configuration. The analysis
of the specific heat at low temperatures yields an entropy of $R\ln8$ around 6~K consistent with a
fully degenerate $J=7/2$ multiplet. We could identify four anomalies in specific heat and
dc-susceptibility evidencing a complex magnetic phase-diagram. At the lowest obtained temperature
($T\approx0.5$~K) we do not observe any saturation of the magnetic moment in fields up to 7~T.
NdPt$_4$Ge$_{12}$ shows one additional magnetic transition inside the magnetically ordered state
which is possibly related to a re-arrangement of the magnetic moments. The analysis of the magnetic
entropy points to a quartet CEF ground state. To clarify the nature of the magnetic order in
EuPt$_4$Ge$_{12}$ and NdPt$_4$Ge$_{12}$ further studies on single crystals would be desirable.

%\section*{Acknowledgements}


\section*{References}
\begin{thebibliography}{99}

\bibitem{Sales1996} Sales B C, Mandrus D and Williams R K (1996) {\it Science} {\bf 272} 1325.

\bibitem{Uher2001} Uher C (2001) {\it Semicond, Semimetals} {\bf 69} 139.

\bibitem{Chen2003} Chen G, Dresselhaus M S, Dresselhaus G, Fleurial J-P and Caillat T (2003) {\it Int.
Materials Rev.} {\bf 48} 45.

\bibitem{Uher2008} Uher C in {\it Chemistry, Physics, and Materials Science of Thermoelectric Materials:
beyond Bismuth Telluride}, eds. Kanatzidis M G, Hogan T P and Mahanti S D, Spinger 2008.

\bibitem{Sales2007} Sales B C 2007 {\it Int. J. Appl. Ceram. Technol} {\bf 4} 291.

\bibitem{Jeitschko1977} Jeitschko W and Braun D (1977) {\it Acta Cryst B} {\bf33} 3401.

\bibitem{Bauer2007} Bauer E {\it et al.} %, A. Grytsiv, X.-Q. Chen, N. Melnychenko-Koblyuk, G. Hilscher, H.
%Kaldarar, H. Michor, E. Royanian, G. Giester, M. Rotter, R. Podloucky, P. Rogl,
(2007) {\it Phys. Rev. Lett.} {\bf99} 217001.

\bibitem{Gumeniuk2008}  Gumeniuk R, Schnelle W, Rosner H, Nicklas M, Leithe-Jasper A and Grin Yu (2008) {\it Phys.
Rev. Lett.} {\bf100} 017002.

\bibitem{Kaczorowski2008} Kaczorowski D and Tran V H (2008) {\it Phys. Rev. B} {\bf 77} 180504R.

\bibitem{Bauer2008} Bauer E {\it et al.} %, X.-Q. Chen, P. Rogl, G. Hilscher, H. Michor, E. Royanian, R. Podloucky, G.
%Giester, O. Sologub, A. P. Gonçalves,
(2008) {\it Phys. Rev. B} {\bf 78} 064516.

\bibitem{Maisuradze2009} Maisuradze A {\it et al.} %, M. Nicklas, R. Gumeniuk, C. Baines, W. Schnelle, H. Rosner, A.
%Leithe-Jasper, Y. Grin, R.  Khasanov,
(2009) {\it Phys. Rev. Lett.} {\bf103} 147002.

\bibitem{Rosner2009} Rosner H {\it et al.} %, J. Gegener, D. Regesch, W. Schnelle,R. Gumeniuk, A. Leithe-Jasper,
%H. Fujiwara, T. Haupricht, T. C. Koethe, H. -H. Hsieh, H. -J. Lin, C. T. Chen, A. Ormeci, Yu. Grin,
%L. -H. Tjeng,
(2009) {\it Phys. Rev. B} {\bf 80} 075113.

\bibitem{Toda2008} Toda M {\it et al.} (2008) {\it J. Phys. Soc. Jpn.} {\bf 77} 124702.

\bibitem{Grytsiv2008} Grytsiv A {\it et al.} (2008) {\it J. Phys. Soc. Jpn. Suppl. A} {\bf 77} 121.


\end{thebibliography}
\end{document}